\documentclass{article}
\usepackage{spconf,amsmath,graphicx}

\title{Exploring Efficient-tuning Methods in Self-supervised Speech Models}

\name{\centering Zih-Ching Chen$^{\star {1}}$, Chin-Lun Fu$^{\star {1}}$, Chih-Ying Liu$^{1}$, Shang-Wen (Daniel) Li$^{2}$, Hung-yi Lee$^{1}$}

\address{$^{1}$ National Taiwan University\\
$^{2}$ Amazon AI}
\copyrightnotice{978-1-6654-7189-3/22/\$31.00~\copyright2023 IEEE}
\begin{document}
%
\maketitle

\begin{abstract}
In this study, we aim to explore efficient tuning methods for speech self-supervised learning. 
Recent studies show that self-supervised learning (SSL) can learn powerful representations for different speech tasks. 
However, fine-tuning pre-trained models for each downstream task is parameter-inefficient since SSL models are notoriously large with millions of parameters. 
Adapters are lightweight modules commonly used in NLP to solve this problem. 
In downstream tasks, the parameters of SSL models are frozen, and only the adapters are trained. Given the lack of studies generally exploring the effectiveness of adapters for self-supervised speech tasks, we intend to fill this gap by adding various adapter modules in pre-trained speech SSL models.
We show that the performance parity can be achieved with over 90\% parameter reduction, and discussed the pros and cons of efficient tuning techniques. This is the first comprehensive investigation of various adapter types across speech tasks.

\end{abstract}
\section{Introduction}
\label{sec:intro}

\begin{figure}[h]
      \centering
      \includegraphics[width=0.47\textwidth]{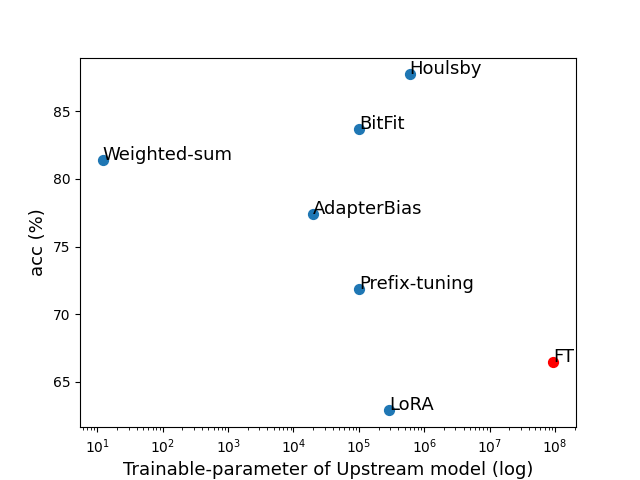}
      \caption{The trade-off between accuracy and number of trained task-specific parameters, for several efficient tuning methods and fine-tuning. The x-axis represents trainable parameter of the upstream model, while the y-axis represents the accuracy of Speaker Identification task (SID). The red point is fine-tuning (FT), and the blue points are the efficient methods.
      }
\label{fg:overview}
\end{figure}

Recently, self-supervised learning (SSL) has gained popularity in the field of computer vision (CV), natural language processing (NLP), as well as speech tasks. SSL pre-trains a shared representation model on a huge amount of unlabeled data. The pre-trained SSL model can be used for various downstream tasks with minimal adaptation via either fine-tuning or utilizing the learned representation from the frozen model \cite{Mohamed2022SSLspeech}. Applying a SSL model to different downstream tasks can significantly lower the entry barrier for developing a model compared to training the model from scratch. Yielding state-of-the-art (SOTA) performance, SSL is desirable for deep learning not only for its outstanding performance, but also for its generalizability and reusability for different tasks in various application scenarios. Transfering from pre-trained models yields strong performance on not only many NLP tasks but speech tasks as well. 

Despite the huge success and popularity SSL has gained, there are some drawbacks when utilizing SSL models. In the presence of various downstream tasks, fine-tuning pre-trained models for each downstream task is still parameter-inefficient since massively self-supervised pre-trained models are notoriously deep, requiring millions or even billions of parameters. Due to this reason, adapting the SSL speech model by fine-tuning requires large storage space. For example, HuBERT X-Large~\cite{hsu2021hubert} contains 964M parameters.
This results in requiring large storage space for each complete set of tuned parameters per downstream task. 
Furthermore, overwriting the pre-trained model parameters may not be the best way of utilizing the pre-trained knowledge from the SSL model.

To overcome these shortcomings, researchers then utilize the SSL speech model by only using the frozen representation~\cite{yang2021superb}. In NLP, efficient tuning techniques have been proposed for leveraging SSL models. One of the most popular efficient methods is adapters~\cite{houlsby2019parameter}, which introduce extra tunable weights and freeze the original parameters of the pre-trained language model (PLM). Adapters have demonstrated comparable performance with fully fine-tuning the entire model while being parameter-efficient. More recently, the prompting technique has shown to be surprisingly effective on PLM~\cite{li2021prefix}. Both methods shows that “freezing” pre-trained models is appealing, especially as model size continues to increase. Rather than requiring a separate copy of the model for each downstream task, a single generalized upstream model can simultaneously transfer to many different tasks. Adapters have been shown to work well for machine translation~\cite{zhu2021counter}, cross-lingual transfer~\cite{pfeiffer2020mad}, as well as transfer learning in automatic speech recognition (ASR)~\cite{thomas2022efficient}. However, these efficient tuning methods are not systematically studied with SSL speech models.

In order to utilize efficient tuning methods to the field of SSL speech representation, in this work, we explore the effectiveness of efficient tuning methods for self-supervised speech models on the SUPERB benchmark~\cite{yang2021superb}. We apply different efficient tuning methods, including adapter tuning and prompt tuning, on SSL speech models with different training objectives. We propose an adapter framework for multiple downstream speech processing tasks, including the recognition tasks, classification, as well as speaker tasks. 
To investigate the effectiveness of these efficient methods, we conduct experiment on 3 SSL models with different training objectives: HuBERT, Wav2vec2~\cite{baevski2020wav2vec}, and DeCoAR2~\cite{ling2020decoar}. The main concept of our work is shown in Fig 1. To our best knowledge, this is the first comprehensive investigation of various efficient tuning methods on different speech tasks. We show that the performance parity can be achieved with over 90\% parameter reduction. Furthermore, we show the pros and cons of various efficient tuning techniques, e.g., the Houlsby adapter~\cite{houlsby2019parameter} is the most efficient in the trade of between performance and the number of parameters, and weighted sum is a very suitable efficient method to use in SSL speech tasks.

\begin{figure}[h]
      \centering
      \includegraphics[width=0.43\textwidth]{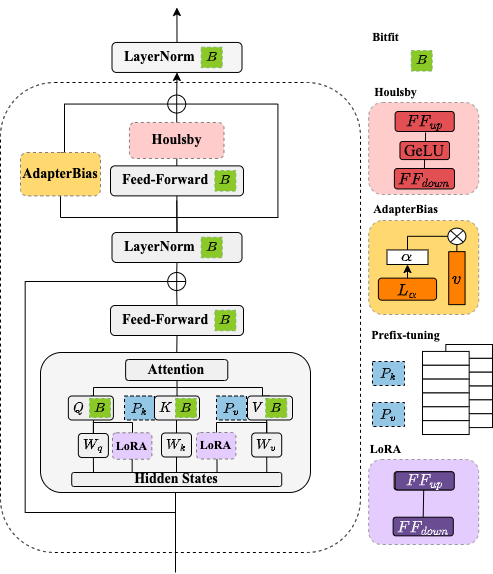}
      \caption{
        Illustration of the transformer architecture and parameter-efficient tuning methods. The blocks with dashed borderlines are the added parameters by the efficient method. $W_q, W_k, W_v$ represents the weights of query, key and value, respectively.
      }
\label{fg:overview}
\end{figure}

\
\section{Related Works}
\label{sec:related}
\vspace{-5pt}

\subsection{Adapter Approach} 
For NLP tasks, adapters are introduced for the transformer architecture. An adapter typically comes with a two-layer feed-forward bottleneck architecture~\cite{houlsby2019parameter}. It was found that adapters approach the performance of full fine-tuning with only a fraction of the parameters in NLP tasks using a PLM. 
Inspired by the success of prompting methods that control PLMs through textual prompts~\cite{brown2020language}, prefix tuning \cite{li2021prefix}, and neural reprogramming \cite{yang2021voice2series} prepends an additional tunable prefix tokens to the hidden layers and only optimized these soft prompts when fine-tuning. More recently, LoRA~\cite{hu2021lora} learns low-rank matrices for parameter updates approximation. AdapterBias~\cite{fu-etal-2022-adapterbias} adds a token-dependent parameter shift to transfer from PLM in a more parameter-efficient manner. Beyond its parameter efficiency, adapter tuning is also shown to be more robust due to its ability to preserve the pre-trained knowledge~\cite{he2021effectiveness}, and often exhibits robustness in out-of-distribution evaluation\cite{li2021prefix}. 

In the field of speech processing tasks, adapters have also been utilized for efficient SSL tuning. Using adapters on Wav2vec2 for efficient tuning for ASR has been proposed~\cite{thomas2022efficient}. Moreover, The work~\cite{fan2022draft} proposes residual adapters (RAs) which are inserted in the pre-trained model to learn domain-related information with the same SSL loss as the pretraining stage. Adapters have also been employed for efficient SSL speech pre-training of new tasks in a continual learning setting~\cite{kessler2022adapter}.  As for prompting, it has been applied to speech task~\cite{chang2022exploration} with a prompt tuning paradigm for Generative Spoken Language Model~\cite{lakhotia2021generative}.

However, the above works either apply adapters on one SSL speech model on a specific task, or they did not examine the different efficient tuning methods on different downstream tasks in a comprehensive way. This leaves the question of whether the efficient tuning methods in NLP will yield the same effectiveness when utilized in speech processing tasks. We hypothesize that we will see the same benefits of adapters in a speech model as in an NLP model, namely parameter efficient transfer of the pre-trained network to different downstream tasks with little performance degradation.



\subsection{The SUPERB benchmark}
As more powerful SSL models are being proposed with more promising performance on various tasks, researchers continually try to find extensive evaluation methods to assess model performance, in the hope of understanding the capability of the learned representation in these models. SUPERB~\cite{yang2021superb} is a framework to benchmark SSL models on 10 speech tasks by learning task-specific predictions heads on top of the frozen shared SSL models. In the SUPERB benchmark, they freeze the upstream SSL model, and learn the downstream model according to the downstream task. During tuning, weighted-sum is applied to learn the optimum representation for the specific downstream task. However, they did not explore the capability of the upstream model with fine-tuning. 
\begin{table*}[h]
\vspace*{-5pt}
\centering
\setlength{\tabcolsep}{5mm}{
\begin{tabular}{c|c|ccccccc}

\hline
Method      & Params & ASR  & PR   & SD   & SID   & SF    & IC    & KS    \\ \hline
FT          & 94.7M  & 6.35 & \textbf{2.45} & 9.32 & 66.48 & 84.87     & 99.10 & 95.87    \\
Baseline    & 0      & 7.09 & 7.74 & 7.05 & 64.78 & 86.25 & 96.39 & 95.32 \\
Houlsby     & 0.60M  & 5.88 & 3.00 & \textbf{4.00} & \textbf{87.71} & 85.87 & \textbf{99.60} & 97.17 \\
AdapterBias & 0.02M  & \textbf{5.54} & 4.19 & 5.48 & 77.38 & 86.60 & 99.50 & 97.30 \\
BitFit      & 0.10M  & 9.34 & 4.23 & 5.13 & 83.68 & \textbf{87.40} & 99.50 & \textbf{97.33} \\
LoRA        & 0.29M  & 6.94 & 8.74 & 7.39 & 62.90 & 86.25 & 96.57 & 96.59  \\
Prefix      & 0.10M  & 6.56    & 4.18    & 8.17    & 71.87     & 85.85     & 99.31     & 97.05     \\
Weighted-sum  & 12     & 6.42 & 5.41 & 5.88 & 81.42 & 88.53 & 98.34 & 96.30 \\ \hline
\end{tabular}}
\caption{Performance of different efficient methods in the SUPERB benchmark. The second column represents additional trainable parameter used in upstream model. Note that except for the "Weight-sum" method, other methods directly use the last layer representation of upstream model as the input of the downstream model.}
\label{tab:1}
\end{table*}
\section{Efficient tuning for self-supervised speech models}
\label{sec:methods}

In this paper, we propose a framework to consistently evaluate the efficient tuning methods for SSL speech models. The framework is designed based on three aspects of the experiment: generalizability, coverage, and comparability.

\subsection{Generalizability}
For the purpose of examining the generalizability of the efficient tuning methods in SSL speech models, this framework includes multiple downstream speech processing tasks, involving the recognition tasks, classification tasks, as well as speaker tasks. For recognition tasks, we examine automatic speech recognition (ASR) and phoneme recognition (PR); classification tasks include keyword spotting (KS), slot filling (SF), and intent classification (IC); and for the speaker tasks, we have speaker identification (SID) and speaker diarization (SD). As for the upstream model, we conduct experiments with different training objectives SSL models: HuBERT, Wav2vec2, and DeCoAR2. The former two models are discriminative models, while DeCoAR2 is a generative model.

\subsection{Efficient tuning approaches}
As for coverage, we implement mainstream efficient tuning methods in NLP, and conduct experiments to understand different efficient methods, as well as their integration with SSL model.

The structure of our framework is shown in Fig~\ref{fg:overview}. In our experiments, we apply adapters at the place where they originally added in NLP. Based on different tasks, we apply different downstream models (i.e. LSTM module, a linear classifier) on top of the transformer network. A set of adapters and the downstream model are trained per task, and the rest of the network remains frozen.

\subsubsection{Houlsby adapter}
\label{ssec:houlsby}

Houlsby adapters \cite{houlsby2019parameter} are small bottleneck modules consisting of a down-projection ($FF_{down}$), a non-linearity ($GeLU$), and an up-projection ($FF_{up}$), with a skip connection. Here, we add Houlsby adapters to the second feed-forward layers of transformer layers. The fully connected layers are initialized as a near identity function.
\subsubsection{LoRA}

\label{ssec:lora}
LoRA~\cite{hu2021lora} reduces the number of trainable parameters by learning pairs of rank-decomposition matrices ($FF_{down}$, $FF_{up}$) while freezing the original weights. This reduces the number of parameters for large language models when adapted to specific tasks. In our work, LoRA is added to the attention modules of transformer layers.

\subsubsection{AdapterBias}
\label{ssec:adapterbias}
AdapterBias~\cite{fu-etal-2022-adapterbias} adds frame-dependent biases to the representation shifts by using a vector ($v$) and a linear layer (${L_\alpha}$). $v$ represents the task-specific shift, and ${L_{\alpha}}$ produces the weights ($\alpha$) for input frames. Thus, with the vector and the weights, AdapterBias can add a frame-dependent shift to the transformer layer. We add AdapterBias module to the second feed-forward layers of transformer layers.

\subsubsection{BitFit}
\label{ssec:bitfit}
Instead of adding additional parameters for adaptation, Bitfit~\cite{zaken2021bitfit} tunes the bias term of each module. In our method, we tune the weight of all modules in the upstream model, such as HuBERT, Wav2vec2, and DeCoAR2.

\subsubsection{Prefix tuning}
\label{ssec:prompt}

For prompt tuning~\cite{li2021prefix} in our unified efficient tuning settings, we use prefix tuning, which could be considered as a variant of adapter \cite{he2022towards}. $l$ trainable prefix vectors were prepended to the multi-head attention modules of all transformer layers. To be more specific, the original key ($K$) and value ($V$) are concatenated with trainable prefix vectors $P_k, P_v \in R^{l \times d}$, where $d$ is the model dimension.
During training, only the prefix vectors and the downstream model are updated, while the upstream model remains fixed. 

\subsubsection{Weighted sum}
In the framework of \cite{yang2021superb}, they weighted the sum of multiple hidden states from the upstream model as the final representation.
In our framework, we regard the weighted-sum technique as an efficient method.


\subsection{Comparability}
For the purpose of the comparability of our proposed framework, we design our downstream model to be similar to the SUPERB benchmark, so that our approach is reproducible and comparable. The configuration setting and the hyper-parameter search is consistent with the SUPERB benchmark so that the efficient tuning methods could be evaluated from the aspect of performance, parameter efficiency, as well as stability, and understand the pros and cons of each method for SSL speech processing tasks. 

Inspired by the SUPERB benchmark, we design our framework to keep the downstream models and their fine-tuning simple, while ensuring the performance across pre-trained models with different efficient tuning methods is comparable. PR, KS, SID, and IC are simple tasks that are solvable with linear downstream models. Hence, we use a frame-wise linear transformation for PR with CTC loss~\cite{graves2006connectionist}; mean-pooling followed by a linear transformation with cross-entropy loss for utterance-level tasks (KS, SID, and IC).
For ASR, a vanilla 2-layer 1024-unit BLSTM is adopted and optimized by CTC loss on characters. The trained model is decoded with LibriSpeech~\cite{panayotov2015librispeech}. 
Regarding SF, slot-type labels are represented as special tokens to wrap the slot values in transcriptions. Similar to the SUPERB benchmark, SF is also re-formulated as an ASR problem. As for SD, we train SD with permutation-invariant training (PIT) loss, which is also used in the SUPERB benchmark.

\section{Experiment}
\label{sec:exp}



\begin{table*}[h]
\centering
\scalebox{1}{
\setlength{\tabcolsep}{1.3mm}{
\begin{tabular}{c|ccc|ccc|ccc}
\hline
            & \multicolumn{3}{c|}{ASR}    & \multicolumn{3}{c|}{SD}     & \multicolumn{3}{c}{KS}      \\ \cline{2-10}
      Method      & HuBERT & DeCoAR2 & Wav2vec2 & HuBERT & DeCoAR2 & Wav2vec2 & HuBERT & DeCoAR2 & Wav2vec2 \\ \hline\hline
FT          & 6.35   & 25.46   &       6.01   & 9.32   & 12.67   & 12.38    & 95.81  & 27.36   & 97.50    \\
Baseline    & 7.09   & 39.06   &  10.79        & 7.05   & 9.14    & 8.07     & 95.32  & 91.69   & 91.95    \\
Houlsby     & 5.88   & 28.63   &    5.99      & \textbf{4.00}   & \textbf{5.76}    & \textbf{4.23}     & 97.17  & 96.07   & 96.75    \\
AdapterBias & \textbf{5.54}   & 29.89   &    \textbf{5.96}      & 5.48   & 6.95    & 6.16     & 97.30  & 96.23   & 91.56    \\
BitFit      & 9.34   & 29.40   &   6.01       & 5.13   & 7.05    & 5.29     & \textbf{97.33}  & \textbf{96.72}   & \textbf{96.85}    \\
LoRA        & 6.94   & 39.52   &    11.32      & 7.39   & 8.78    & 7.92     & 96.59   & 25.38    & 92.80     \\
Prefix      &   6.56    &    \textbf{13.48}     &    6.54      &    8.17    &   10.05      & 10.01          &    97.05    &    88.7     &     96.82     \\
Weighted-sum  & 6.42   & 36.26   &    6.43      & 5.88   & 5.88    & 6.08     & 96.30  & 94.48   & 96.23  \\ \hline
\end{tabular}}
}

\caption{Performance of different upstream models. We used three different objective self supervise speech models: HuBERT, DeCoAR2, and Wav2vec2.}
\label{tab:2}
\end{table*}
\subsection{Performance on the SUPERB benchmark}
\label{exp:1}
We explore different efficient methods in the SUPERB benchmark. Note that 'FT' represents fine-tuning. The 'Baseline' here means that we tune the downstream model only. The tasks we have examined can be categorized into three: recognition task, classification task, and speaker task. The result is shown in Table~\ref{tab:1}. In general, most efficient methods perform better than Baseline and FT. For the classification tasks (i.e. KS, IC), Baseline already yields good performance. Thus, the improvement in using efficient methods is not apparent. For recognition and speaker tasks (i.e. ASR, PR, SD, SID), the advantage of using efficient methods can be seen. Especially in SID, Houlsby improves 23\% accuracy compared to Baseline.
On average, Houlsby yields high and stable performance among all efficient methods since it has the biggest trainable parameter. For LoRA, it performs worst among efficient methods and even worse than Baseline in some tasks (i.e. PR, SD, SID). 
One thing worth mentioning is that Weighted-sum is a powerful and efficient method for speech tasks, where it gets comparable performances in the SUPERB benchmark by just adding a few trainable parameters to the upstream model.

\newcommand\graphwidth{0.42}
\begin{figure}
      \centering
      \includegraphics[width=\graphwidth\textwidth]{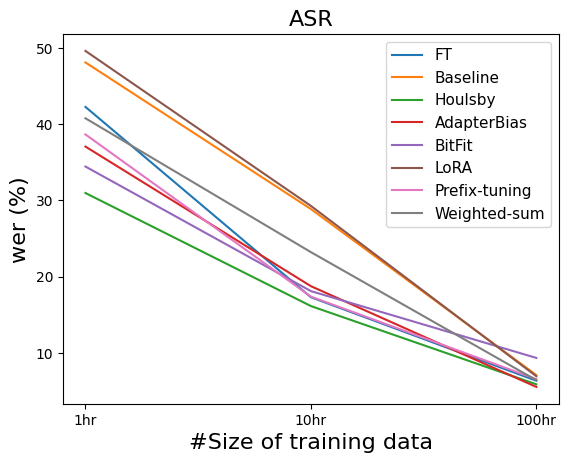}
      \hfill
      \includegraphics[width=\graphwidth\textwidth]{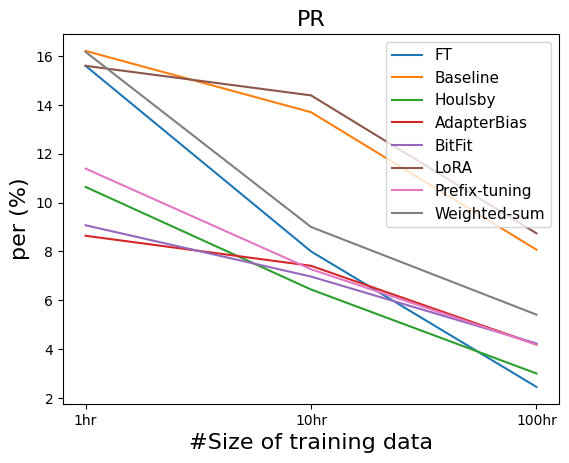}
      \caption{
        Performance of efficient methods in low-resource adaptation. We train our model on 1-hour data and 10-hour data from Libri-Light and test the model on LibriSpeech. The x-axis represents the size of training data, while the y-axis represents performance of each task. For ASR, we report word error rate (WER). For PR, we report phone error rate (PER).}
\label{fg:ASR}
\end{figure}

\subsection{Upstream models with different training objectives}
\label{exp:2}
We also examine the generalization ability of these efficient methods with upstream SSL speech models with different training objectives. We use three different training objective models as upstream models: HuBERT, DeCoAR2, and Wav2vec2. As shown in Table~\ref{tab:2}, efficient methods all gain comparable performance when applied to different upstream models. For example, in SD, Houlsby performs best when using HuBERT, DeCoAR2, and Wav2vec2; in KS, BitFit performs best. 

Moreover, the improvement of utilizing efficient methods depends on the upstream model. If the upstream model already yields strong performance in Baseline, the performance gain of using efficient methods becomes less. In contrast, if Baseline does not get a strong performance, the improvement of using efficient methods is more significant. For ASR, we can observe that Houlsby adapter improves 1.21\% word error rate (WER) than Baseline when the upstream model is HuBERT. However, when the upstream model is DeCoAR2, using Houlsby adapter improves 10.43\% WER.


\subsection{Low-resource Adaptation}
\label{exp:3}

In NLP, adapters are shown to have advantages over fine-tuning when adapting to low-resource datasets~\cite{he2021effectiveness, zaken2021bitfit, fu-etal-2022-adapterbias}. To see if this property also holds when applied in speech tasks, we trained different efficient methods in the low-resource settings. All methods were trained with 1-hour and 10-hour datasets generated by Libri-Light and tested on the testing set of LibriSpeech. We conducted experiments on recognition tasks, including ASR and PR. As shown in Fig 3, the efficient methods perform better than fine-tuning in the low-resource settings. We observed a similar tendency in speech tasks. As the training data becomes smaller, tuning the majority of the parameters may result in a higher risk of overfitting the training data. Using adapter methods helps overcome this issue. Also, we found that LoRA failed to achieve comparable performance in the low resource settings as it cannot perform well in speech tasks generally. For PR, fine-tuning performs better than Houlsby adapter in 100-hour training data. However, as the size of training data decreases, the benefit of efficient tuning methods started to emerge. As shown in Fig 3, in 10-hour and 1-hour, Houlsby adapter started to perform better than fine-tuning.

\subsection{Analysis} 
\label{exp:4}
In this part, we explore the benefit of efficient tuning methods beyond parameter-efficiency from two aspects: stability and learning rate robustness.

\begin{table}[t]
\centering
\scalebox{0.8}{
\begin{tabular}{c|cc|cc}
\hline
            & \multicolumn{2}{c|}{ASR}                      & \multicolumn{2}{c}{PR}         \\ \cline{2-5} 
Method      & \multicolumn{1}{c|}{1hr}        & 10hr        & \multicolumn{1}{c|}{1hr} & 10hr \\ \hline
FT          & \multicolumn{1}{c|}{35.38\footnotesize$\pm$\textbf{6.02}} & 15.30\footnotesize$\pm$1.73  & \multicolumn{1}{c|}{15.60\footnotesize$\pm$\textbf{5.56}}   & 6.15\footnotesize$\pm$0.98    \\
Baseline    & \multicolumn{1}{c|}{51.54\footnotesize$\pm$3.04} & 24.51\footnotesize$\pm$3.85  & \multicolumn{1}{c|}{16.90\footnotesize$\pm$0.63}   & 12.93\footnotesize$\pm$0.72    \\
Houlsby     & \multicolumn{1}{c|}{31.67\footnotesize$\pm$0.94} & 16.67\footnotesize$\pm$4.20  & \multicolumn{1}{c|}{11.26\footnotesize$\pm$0.62}   & 6.44\footnotesize$\pm$0.13    \\
AdapterBias & \multicolumn{1}{c|}{37.97\footnotesize$\pm$1.45} & 19.35\footnotesize$\pm$5.92  & \multicolumn{1}{c|}{8.83\footnotesize$\pm$0.32}   & 6.96\footnotesize$\pm$0.39    \\
BitFit      & \multicolumn{1}{c|}{35.47\footnotesize$\pm$1.39} & 15.17\footnotesize$\pm$2.54 & \multicolumn{1}{c|}{8.89\footnotesize$\pm$0.34]}   & 6.93\footnotesize$\pm$0.09    \\
LoRA        & \multicolumn{1}{c|}{51.27\footnotesize$\pm$1.62} & 29.47\footnotesize$\pm$\textbf{8.20}  & \multicolumn{1}{c|}{15.96\footnotesize$\pm$0.32}   & 14.85\footnotesize$\pm$\textbf{1.64}    \\
Prefix      & \multicolumn{1}{c|}{39.00\footnotesize$\pm$1.17}          & 17.71\footnotesize$\pm$1.04           & \multicolumn{1}{c|}{11.75\footnotesize$\pm$0.35}   & 7.25\footnotesize$\pm$0.08    \\
Weighted-sum  & \multicolumn{1}{c|}{43.84\footnotesize$\pm$2.15} & 23.22\footnotesize$\pm$7.77  & \multicolumn{1}{c|}{13.58\footnotesize$\pm$2.24}   & 9.04\footnotesize$\pm$0.08    \\ \hline
\end{tabular}}
\caption{Performance of different low-resource data in efficient methods. We train with three random seeds and report the mean and standard deviation.}
\label{tab:3}
\end{table}
\begin{table}[t]
\centering
\scalebox{0.85}{
\begin{tabular}{c|c|c|c|c}
\hline
Method      & 5$\times$$10^{-6}$      & 5$\times$$10^{-5}$      & 5$\times$$10^{-4}$      & 5$\times$$10^{-3}$      \\ \hline
FT          & 3.03$\footnotesize$$\pm$0.1  & 2.81$\footnotesize$$\pm$0.4  & 100$\footnotesize$$\pm$0     & 100$\footnotesize$$\pm$0     \\
Houlsby     & 6.09$\footnotesize$$\pm$0.49 & 3.24$\footnotesize$$\pm$0.14 & 2.81$\footnotesize$$\pm$0.03 & 3.06$\footnotesize$$\pm$0.03 \\
AdapterBias & 7.54$\footnotesize$$\pm$0.06 & 4.52$\footnotesize$$\pm$0.01 & 3.79$\footnotesize$$\pm$0.02 & 3.72$\footnotesize$$\pm$0.02 \\ \hline
\end{tabular}}
\caption{Performance of different methods with different learning rates. The downstream task is PR. We run 5 different random seeds and report the mean and standard deviation.}
\label{tab:robustness}
\end{table}
\subsubsection{The stability of low-resource adaptation}

In this section, we use the Libri-Light tool to split different low-resource data from LibriSpeech with different random seeds. For each efficient method, we run three random seeds and compute the mean and standard deviation. From Table 3, we can find that efficient methods have more tolerant than FT when the training data becomes less. Compared with ASR and PR, ASR has a bigger standard deviation than PR. The reason may be that we use a more complex downstream model (2 layers of LSTM) in ASR. Training with low-resource data would make the complex model more unstable than a simple downstream model (i.e. a linear layer) used in PR.

\subsubsection{Learning rate robustness of efficient tuning methods}

This part evaluates the tolerance of the learning rate in different methods. Here we pick fine-tuning (FT), Houlsby adapter, and AdapterBias since Houlsby adapter has the biggest trainable parameters and AdapterBias has the lowest parameters. In Table 4, we train on PR and learning rates ranging from 5$\times$$10^{-6}$ to 5$\times$$10^{-2}$. We can observe that FT has less tolerance than efficient methods. FT does not work on larger learning rates, while efficient methods receive more stable performance among a large range of learning rates. Comparing with Houslby adapter and AdapterBias, AdapterBias has smaller standard deviation than Houlsby adapter since AdapterBias has less trainable parameters than those of Houlsby adapter. Thus, with less trainable parameters, the model would not overfit to training data.


\subsection{Discussions}
\label{exp:5}
In this section, we discuss the strength and limitation of efficient tuning methods in speech processing tasks, as well as their behavioral difference from NLP.

\vspace{-5pt}
\subsubsection{Performance analysis of adapter methods}
From the experimental results, we found that Houlsby adapter performs the best among all efficient tuning methods. This is different from NLP, as in NLP, the overall performance gain of Houlsby adapter is not that significant~\cite{he2022towards}. In the SUPERB benchmark, Houlsby adapter outperforms other efficient methods in 3 out of 7 tasks.

LoRA is an effective adapter in NLP, achieving comparable performance with other adapters~\cite{hu2021lora}. However, it performs worst in the SUPERB benchmark. We guess that the position added adapters play a crucial role. Both Houlsby adapter and AdapterBias are added behind the second feed-forward layer, while LoRA is added in the attention module. Therefore, in SUPERB benchmark, adding adapters in the feed-forward layer is more effective than adding adapters in the attention module.

In NLP, prefix-tuning achieves comparable performance with adapter methods~\cite{he2022towards}. Nonetheless, prefix-tuning does not perform better than adapter methods in the SUPERB benchmark. One reason may be the initialization of prefix-tuning significantly affects the performance in speech tasks. The embedding is discrete in NLP tasks, while in speech tasks, each frame representation is continuous. Thus, we initialize the prefix with the average of the hidden states of the first batch of data. However, it is still worth designing a suitable initialization of prompt in the future.

In addition, weighted-sum is not a common technique in NLP. Nevertheless, weighted-sum improves a huge performance in the SUPERB benchmark. In the work \cite{pasad2021layer}, they find that output from each layer of speech SSL model contain information related to different tasks. Therefore, weighted-sum leverages information from different layers and receives high performance in speech tasks.

\vspace{-1pt}
\subsubsection{Performance analysis of different types of tasks}
In NLP, most efficient methods work well on classification tasks, but do not perform as well in generative tasks. In the SUPERB benchmark, utilizing efficient methods achieves good performance in general on not only classification tasks (i.e. IC, KS), but also generative tasks, such as ASR.
However, there are some tasks (i.e. PR, SF) where efficient methods do not work very well.
In the future, it is worth designing a suitable adapter for speech and considering more challenging tasks, such as Out-of-domain Automatic Speech Recognition Tasks (OOD-ASR).

\vspace{-4pt}
\section{Conclusion}
\label{sec:conclusion}

In this paper, we explore the effectiveness of efficient tuning methods for SSL speech representation transfer. We proposed a framework to consistently evaluate efficient tuning methods for SSL speech models. Extensive experiments are conducted to investigate the various adapter types on different SSL speech models on a wide range of speech processing tasks. Other than finding adapters capable of achieving comparable performance to the fully fine-tuned models, we further examine the stability of adapters compared with fine-tuning. We then discussed on comparing efficient methods in NLP and Speech. To our best knowledge, this is the most comprehensive work exploring adapter methods on a wide range of downstream speech tasks so far.

\section{ACKNOWLEDGMENTS}
\label{sec:ack}
\vspace{-5pt}
We thank Taiwan Web Service Corporation and National Center for High-performance Computing (NCHC) of National Applied Research
Laboratories (NARLabs) in Taiwan for providing computational and storage resources. Also, part of the work presented here was carried out during the 2022 Jelinek Memorial Summer Workshop on Speech and Language Technologies at Johns Hopkins University, which was supported with unrestricted gifts from Amazon, Microsoft, and Google.

\bibliographystyle{IEEEbib}
\bibliography{strings,refs}

\end{document}